\documentclass[12pt]{article}

\usepackage{amsmath,amssymb}
\usepackage{graphicx}

\let\eps=\varepsilon

\begin{document}



\begin{center}

{\Large{Dynamics of Holstein polaron in a chain
\\
with thermal fluctuations}}
\\

\vspace{5pt}
N.S.~Fialko, V.D.~Lakhno 

\vspace{5pt}

{\small{Institute of Mathematical Problems of Biology RAS --
the Branch of Keldysh Institute of Applied Mathematics of Russian Academy of Sciences,

1, Prof.~Vitkevich street, 142290, Pushchino, Moscow region, Russia.
\\
E-mail: fialka@impb.ru}}
\end{center}

\begin{abstract}
Numerical modeling is used to investigate the dynamics of a polaron in a chain
with small random Langevin-like perturbations which imitate the environmental
temperature $T$ and under the influence of a constant electric field.
In the semiclassical Holstein model the region of existence of polarons
in the thermodynamic equilibrium state depends not only on temperature but also on the chain length.
Therefore when we compute dynamics from initial polaron data,
the mean displacement of the charge mass center differs for different-length chains at the same temperature.
For a large radius polaron,
it is shown numerically that the ``mean polaron displacement'' (which takes account only
of the polaron peak and its position) behaves similarly for different-length chains
during the time when the polaron persists. A similar slope of the polaron displacement
enables one to
find the polaron mean velocity and, by analogy with the charge mobility, assess the ``polaron mobility''.
The calculated values of the polaron mobility for $T \approx 0$ are close to the value at $T=0$,
which is small but not zero.
For the parameters corresponding to the small radius polaron,
simulations of dynamics demonstrate switching mode between immobile polaron and delocalized state.
The position of the new polaron is not related to the position of the previous one;
charge transfer occurs in the delocalized state.
\end{abstract}

\noindent
\texttt{Keywords:} Langevin  thermostat, polaron mobility, small radius polaron, large radius polaron, homogeneous DNA.
\\
\texttt{2000 MSC:}
34C60, 
60J65 


\section{Introduction}

Discrete nonlinear systems have appeared in the various fields
of physics, chemistry and biology \cite{bib01,bib1,bib02,bib03,bib04,bib05,bib06,bib07,bib2}
and have attracted considerable attention recently.
The problem of charge transfer in quasi-one-dimensional molecules, such as DNA or proteins, is of interest to biophysics.
Due to its stability the polaron mechanism of charge transfer attracts attention of a great number of researchers.
Presently
it is believed that current carriers in DNA are polarons or solitons, see for example \cite{bib1,bib2,bib3,bib4} and references therein.
The interest in the charge transfer mechanisms is also associated with the possibility of using DNA in molecular electronics \cite{bib3,bib4,bib5}.

Here we calculate the polaron properties
in the semi-classical Holstein model,
where chain of sites corresponds to the simplest representation of the DNA duplex as a sequence of nucleotide pairs.
The polaron dynamics was studied earlier in the Holstein model for unperturbed chains in an external electric field
both analytically (in a continuous medium) \cite{bib6} and numerically \cite{bib7,bib8,bib9,bib10}.
It was assumed that weak excitations of the chain (with energy much less than the characteristic energy
equal to the depth of the polaron level) change the polaron properties only slightly.

However in the case when the chain is always affected by a random force (Langevin thermostat)
weak excitations influence considerably the polaron characteristics.
In the paper by Lomdahl and Kerr \cite{bib11} direct numerical experiments demonstrated that
Davydov soliton at physiological temperatures quickly decays and cannot transfer energy. For the soliton coupling energy of about 300\,K,
its disruption occurs at low temperature less than 10\,K. Subsequent numerical experiments and analytical estimate of different authors \cite{bib12,bib13,bib14,bib15,bib16} confirmed the conclusion that at physiological temperatures
in biomacromolecules polarons decay and a charge is delocalized.

Earlier for a Holstein model a charge mobility $\mu$ was calculated in the range of ``high'' temperatures $T$ \cite{bib17},
when a charge is delocalized, and the temperature dependence of the mobility is estimated \cite{bib18}.
The dependence $\mu(T) \sim T^{-2.3}$ \cite{bib18} means a growth of the mobility as the temperature decreases.
This estimation is applicable in the range of high temperatures. It is assumed that in the range of low temperatures
a charge forms a polaron state in a chain with far less mobility \cite{bib5,bib19}.

In an unperturbed chain (at $T=0$), the polaron dynamics does not depend on the chain length
(if the chain is much longer than the polaron size, of course).
E.g., under the action of an electric field with constant intensity,
polarons move with the same velocity in the chains of different length.
The results of modeling at $T \neq 0$ demonstrate \cite{bib20,bib21} that
in the thermodynamic equilibrium state (TDE) the existence of polaron depends not only on the thermostat temperature $T$,
but also on the length of the chain $N$, i.e.\ on the thermal energy of the chain $NT$.
Hence, at the same temperature, in the TDE in short chains a polaron does not decay while in long chains
(heat energy of the chain grows with $N$) a charge is delocalized.
In the region of polaron existence, at the same temperature in the TDE
the mean characteristics, such as delocalization parameter and maximum probability of charge localization,
depend on the chain length. Therefore in this region there are no stationary processes
(similar to the root-mean-square deviation $\langle X^2(t)\rangle$ or
mean displacement $\langle X(t)\rangle$ under the influence of an external electric field) which depend only on temperature.

Here we present the results of modeling the charge dynamics in different-length chains
from the initial data ``a polaron\,$+$\,thermal fluctuations of the chain''.
Numerical simulations enabled us to find some general dependencies at the same temperature and
the same intensity of the external electric field
at the first stage when a polaron has not yet disrupted.
Small temperature region is considered
where in the TDE a charge forms a polaron in relatively long chains.
The results obtained are partly in a ``nonphysical region'': for temperatures below Debye temperature $\Theta$,
a classical description of the sites motion is inapplicable \cite{bib11}.
For this reason the simulation results are of qualitative interest.
Two cases are considered -- small radius polaron (SRP) with the parameters partly corresponding to homogeneous adenine DNA chains
and a large radius polaron (LRP) with the parameters of thymine fragments.

Presently, there are many papers devoted to modeling the motion of a charged particle in various-type molecular
chains (e.g., \cite{bib16,addR1,addR2,ad2,ad3,ad4,ad6,ad7,ad9}, also reviews \cite{bib4,ad10} and references therein).
In some papers polaron is considered (according to the physical definition of SRP) for parameter values
at which the polaron is localized at several sites.
Here for SRP we consider the case when the charge is localized at one site with probability of almost 1,
and LRP is localized in the region of about 15 sites.
Also other works do not discuss the effect of chain length on the polaron state at a given temperature,
most likely because with ``physically significant'' parameter values such effect is hard to notice on the computation time intervals.
In our simulation we chose the adapted values of the parameters that speed up the system's movement to the TDE,
and we suggest that not only the mean values in the TDE are the same for two systems with the same
ratio of parameters \cite{bib21}, but the processes of reaching the TDE state from
the same initial polaron state will be qualitatively similar.

The paper is arranged as follows.
In Section \ref{sec2} we describe the model and motion equations, in Section \ref{sec3}
the model parameters and the initial data are given, in Section \ref{sec4} we describe the results of modeling the SRP,
in Section \ref{sec5} the results of LRP calculations are given and
(by analogy with the charge mobility) estimation of the polaron mobility is made.

\section{Model}
\label{sec2}

The model is based on the Holstein Hamiltonian for discrete chain of sites \cite{bib22}.
In the semiclassical approximation, choosing the wave function $\Psi$ in the form
$\Psi = \sum_{n=1}^N b_n |n\rangle$,
where $b_n$ is the amplitude of the probability of the charge
(electron or hole) occurrence at the $n$-th site ($n=1,{\ldots} ,N$, $N$ is the chain length)
the averaged Hamiltonian has the form:
\begin{align}
\langle \Psi | \hat{H} |\Psi \rangle &=
\sum_{m,n} \nu_{nm} b_m b_{n}^* +
\frac 12 \sum_n M \dot{\tilde{u}}_{n}^2 +
\nonumber \\ &{}+
\frac 12 \sum_n K \tilde{u}_{n}^2 +
\sum_n \alpha' \tilde{u}_{n}  b_{n} b_{n}^* +
\sum_n e \tilde{V} a n b_{n} b_{n}^*.
\label{eq1}
\end{align}
Here $\nu_{mn}$ ($m\neq n$) are matrix elements of the charge transition
between $m$-th and $n$-th sites (depending on overlapping integrals),
$\nu_{nn}$ is the electron energy on the $n$-th site.
We consider the nearest neighbour approximation: $\nu_{mn} = 0$, if $m \neq n\pm 1$.
The charge energy at the sites depends linearly on the sites displacements $\tilde{u}_n$,
$\alpha'$ is the coupling constant,
$M$ is the $n$-th site's effective mass, $K$ is the elastic constant.
We deal with homogenous chains and choose $\nu_{nn} = 0$.
The last term in \eqref{eq1} takes account of the constant external field of intensity $\tilde{V}$,
$e$ is the electron charge, $a$ is the distance between neighbouring sites.

Motion equations for Hamiltonian~\eqref{eq1} have the form:
\begin{align}
i \hbar \frac{d b_n}{d \tilde{t}} &=
\nu_{n,n-1} b_{n-1} + \nu_{n,n} b_n +
\nu_{n,n+1} b_{n+1} + \alpha' \tilde{u}_n b_n + ean \tilde{V} b_n,
\label{eq2a} \\
M \frac{d^2 \tilde{u}_n}{d \tilde{t}^2} &=
-K \tilde{u}_n - \alpha' |b_n|^2 -
\tilde{\gamma} \frac{d \tilde{u}_n}{d \tilde{t}}
+ \tilde{A}_n (\tilde{t}).
\label{eq2b}
\end{align}
To model a thermostat, subsystem \eqref{eq2b} involves the terms with friction
($\tilde{\gamma}$ is the friction coefficient) and the random force $\tilde{A}_n (\tilde{t})$ such that
$\langle \tilde{A}_n (\tilde{t}) \rangle =0$,
$\langle \tilde{A}_n (\tilde{t}) \tilde{A}_m (\tilde{t}+\tilde{s}) \rangle =
2 k_B \tilde{T} \tilde{\gamma} \delta_{nm} \delta (\tilde{s})$
($\tilde{T}$ is the temperature [K], $k_B$ is Boltzmann constant).
This way of imitating the environmental temperature with the use of Langevin equations \eqref{eq2b}
is well known \cite{bib11,bib23}.

Motion equations \eqref{eq2a},\eqref{eq2b} for homogeneous chain after nondimensionalization have the form
\begin{align}
i \frac{d b_n}{d t} &=
\eta ( b_{n-1} + b_{n+1} ) + \chi u_n b_n + n V b_n,
\label{eq3a} \\
\frac{d^2 u_n}{d t^2} &=
-\omega^2 u_n - \chi |b_n|^2 - \gamma \frac{d u_n}{d t} + \xi Z_n (t).
\label{eq3b}
\end{align}
The relation between dimension and dimensionless parameters is as follows.
The matrix elements $\eta = \nu_{n,n\pm1} \tau / \hbar$,
the frequency of sites oscillation $\omega = \sqrt{\tau^2 K/M}$
($\tau$ is the characteristic time, $\tilde{t}  = \tau t$).
The coupling constant is $\chi = \alpha' \sqrt{ \tau^3 / \hbar M}$;
$\gamma = \tau \tilde{\gamma} / M$;
$V = ea\tilde{V} \tau/\hbar$.
$Z_n(t)$ is random term with the properties
\begin{align}
\langle Z_n(t) \rangle = 0, \quad
\langle Z_n(t)  Z_m(t+t') \rangle = \delta_{mn}\delta(t'),
\nonumber \\
\xi = 
\sqrt{ \frac{2 k_B T^* \tau}{\hbar}} \sqrt{ \gamma T}, \qquad
\label{temper}
\end{align}
where the dimensionless temperature is $T = \tilde{T}/T^*$, $T^*$ is characteristic value.
Though the chain is homogeneous, under the electric field with intensity $V$
the electron energy on the $n$-th site is $nV$.

For $T=0$, $V=0$ the state with the lowest energy will be a polaron \cite{bib19,bib22};
in this case the velocities of the sites $\dot{u}_n =0$, and displacements are:
\begin{align}
u_n = - \frac{\chi}{\omega^2} |b_n|^2.
\label{eq6}
\end{align}
Theoretical estimates of total energy for Hamiltonian \eqref{eq1} in the case of SRP
when $\eta$ is small and a charge is localized at one ($n$-th) site with the probability $p_n = |b_n|^2 \approx 1$
\begin{align}
E \approx - \frac 12 \frac{\chi^2}{\omega^2},
\label{eq6a}
\end{align}
and in the case of LRP when the probabilities change only slightly at neighboring sites \cite{bib22}
\begin{align}
E \approx 2\eta + \frac 1{48} \frac{\chi^4}{\eta \omega^4}
\label{eq6b}
\end{align}
(here $\eta < 0$).

\section{Setting up a computational experiment}
\label{sec3}

For a predetermined temperature of the thermostat $T$,
we compute a set of realizations (trajectories of the system \eqref{eq3a}, \eqref{eq3b} from different initial data
and with different pseudo-random time-series) and calculate time dependencies averaged over realizations (``by ensemble'').
We consider finite chains with free ends, i.e.\ initial value problem.

For the initial data we use polaron (the amplitudes of probability $b_n$ and displacements \eqref{eq6} corresponds to the lowest energy for $T=0$),
Gaussian random variables are added to these displacements,
and the velocities of the sites $\dot{u}_n$ are determined by Gaussian random variables with a relevant distribution.

\begin{figure}[htb]
\includegraphics
[scale=0.15]{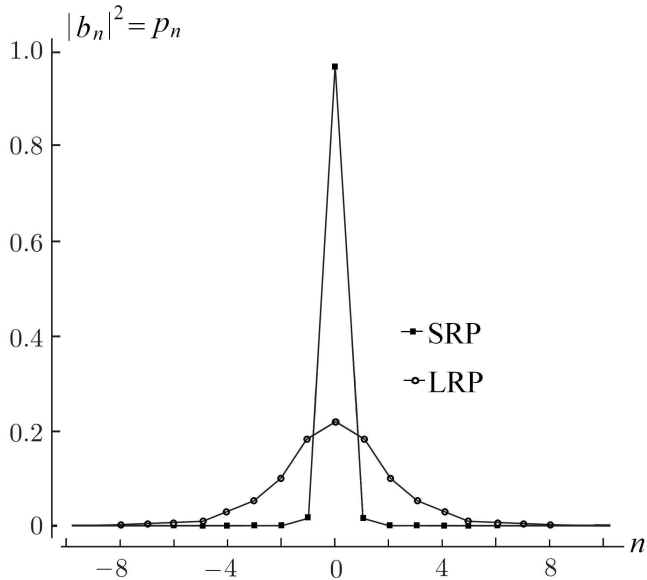}
\caption{Polarons with parameter values for polyA (SRP) and polyT (LRP) at $T=0$.
\label{fig1}}
\end{figure}

The parameters of the model corresponding to nucleotide pairs \cite{bib24,bib25,bib26} are:
$M = 10^{-21}$\,g,
terahertz
frequency of the sites oscillations $\tilde{\omega} = 10^{12}$\,sec$^{-1}$ corresponds to
rigidity of the hydrogen bonds $K \approx 0.062$\,eV/\AA$^2$, the coupling constant $\alpha' = 0.13$\,eV/\AA.
In the adenine chain polyA, for neighboring adenines, the matrix element of the transition between the sites is
$\nu_{\mathrm{AA}} = 0.030$\,eV, and in the thymine fragment polyT $\nu_{\mathrm{TT}} = 0.158$\,eV.
In dimensionless form in choosing the characteristic time $\tau = 10^{-14}$\,sec this corresponds to
$\eta_{\mathrm{AA}} = 0.456$,  $\eta_{\mathrm{TT}} = 2.4$, $\omega =0.01$, $\chi =0.02$ ($\chi/\omega =2$).
The characteristic temperature value is $T^* = 1$\,K.
In the polyA chains SRP is formed, for $T=0$ in the state with the lowest energy
a charge is localized at one site with the probability of $p_n = |b_n|^2 \approx 0.97$;
in polyT chains LRP exists, 
a charge is localized in the region less than 20 sites
with the maximum probability $\max_n p_n = p_M \approx 0.22$.
Fig.\ \ref{fig1} shows the probability distribution for polyA and polyT.

The relations \eqref{eq6a},\eqref{eq6b} suggest that the system energy depends on the ratio $\chi/\omega$.
It was shown \cite{bib21} for homogeneous chains, that in the TDE the mean values
depend not on the parameter values but on their ratios: in systems with the parameters
\begin{align}
\{ \eta, \chi, \omega\} \quad \text{and} \quad \{ \eta, C \chi, C \omega\} \quad (C = const)
\label{eq7}
\end{align}
the averaged distribution of the probabilities, the total energy, the averaged polaron size
$\langle R \rangle$ are similar. However the time to reach the TDE from
the polaron initial state depends on the classical frequency of the system $\omega$.

In numerical simulations for the classical subsystem we mainly used adap\-ted values of the parameters $\omega = 0.5$, $\chi = 1$
($\chi/\omega = 2$), for which the system quicker comes to the TDE.
We suggest that a qualitative picture of polaron decaying is similar for different parameters in which the ratio $\chi/\omega$ holds;
some test calculations do not contradict this assumption.

The possibility of polaron transfer for different intensities $V$ from 0 to 0.1 was investigated.
The value $V = 0.1$ for choosing DNA parameters
corresponds to the intensity $\tilde{V} \approx 3\cdot 10^7$\,V/m. This value is not very large in nanometer scale;
thus, in experiments \cite{bib27} the voltage up to 10\,V was applied to the ends of 20-site DNA chain.
The distance between the neighboring base pairs is $a \approx 0.34$\,nm, i.e.\ $\tilde{V} \approx 3\cdot 10^8$\,V/m.

In all the realizations at $t=0$ the polaron center was localized at one and the same site $n_0$.
In the numerical experiment we calculate (and then for LRP we usually average over 50--100 realizations):
the probabilities to find the charge at the $n$-th site $p_n(t)$,
and the parameter of delocalization $R(t)$
\begin{align}
R = \frac{1}{\sum_1^N |b_n|^4} = \frac{1}{\sum p_n^2}.
\label{eqRv2}
\end{align}
(In other papers \cite{addR1,addR2,addR3,ad3} $R(t)$ is also called participation number, localization length,
degree of delocalization or measure of the locality of a polaron).
This value correlate with charge distribution in the chain.
If a charge is localized at one site, $p_n(t) \approx 1$, then $R(t)\approx 1$.
If a charge is uniformly distributed over a $N$-site chain, $p_n = 1/N$, then $R=N$.
If the charge passes on to the delocalized state at $T \neq 0$, then in the TDE
averaged over realizations $\langle R \rangle \approx N/2$ \cite{bib20}.
Value of $R$ may be associated with polaron radius.
For $T=0$ in polyA chains (SRP) $R \approx 1.07$,
in polyT chains (LRP) $R \approx 6.8$.

Also in the realization the probability maximum $p_M(t) = \max_n p_n(t)$
and $n_M(t)$ -- the number of the site at which $p_M$ is localized -- are calculated
(similar method for soliton recognizing was proposed in \cite{bib11}).
For $V \neq 0$, we calculate the mean displacement of the charge center of mass
\begin{align}
X(t) = \sum_n p_n(t) (n-n_0).
\label{eq8}
\end{align}
Knowing the dependencies  $\langle X(t) \rangle$  for different $V$,
we can estimate the charge velocity $v$ from the slope on a linear fragment (if any)
and assess the charge mobility $\mu$:
\begin{align}
v = \Delta X / \Delta t , \quad \mu = v/V
\label{eq8a}
\end{align}
(theoretically, $\mu$ should be equal for different $V$ for region of the Ohm's linear law).
Similarly we calculate
\begin{align}
X_M = p_M (n_M-n_0)
\label{eq9}
\end{align}
(a certain analog of $X$, taking account of only the probability maximum)
and keep track of the displacements $u_M(t)$ of the sites that have maximum probability.
Besides in the ``chain with window'' where the fragment with the center $n_M$ is not considered
(we usually took out $L = 5$ sites on either side for SRP, i.e.\ 11 sites with the center $n_M$ for a SRP,
and $L = 10$ sites for LRP, i.e.\ 21 sites with the center $n_M$),
we choose a site with the greatest in modulus displacement
\begin{align}
u_K : |u_K| = \max_{|n-n_M|>L} |u_n| .
\label{eq10}
\end{align}
If a charge is in the polaron state, i.e.\ it is localized in a small region,
then outside this region displacements are mainly determined by random fluctuations.
This is an additional check up: if $|u_K| > |u_M|$, then we assume that there is not a polaron in the chain;
if in this case $p_M$ is much larger than the mean probability $\langle p_n\rangle \approx 1/N$ in the chain,
this may be a certain analog of a solectron (a charge was pulled into the well with the largest displacement) \cite{bib29};
however, in simulations we did not observe such a situation.

Calculations of individual realizations for the values $\omega = 0.5$, $\chi = 1$
were performed by 2o2s1g-method \cite{bib30} with forced normalization procedure.
The system (\ref{eq3a},\ref{eq3b}) has the first integral: the total probability
of the charge occurrence in the chain $\Sigma = \sum_n p_n = \sum_n |b_n|^2$ must be equal to 1.
The time intervals of computation can be very long.
As result of numerical integration $\Sigma$ is not kept exactly.
One of the ways that allows us to make computation intervals longer is forced
normalization, when the variables $b_n$ will be ``corrected'' if $|\Sigma - 1|>\eps$:
after the integration step we obtain values $b_n[old]$, calculate $\Sigma[old]$,
and $b_n[new] = b_n[old]/(\Sigma[old])^{1/2}$, so $\Sigma[new] =1$.
By test computations, for integration step $h=0.0005$ (or less)
in the individual trajectory (with the same random seed and initial data)
for $\eps = 10^{-4},10^{-5}$ and $10^{-6}$ the differences
$|(u_n)_{\eps1} - (u_n)_{\eps2}|$ and $|(p_n)_{\eps1} - (p_n)_{\eps2}|$ is less then $10^{-6}$.
Also we calculated the set of 50 realizations for $h=0.0002$
and compared the averaged values in the TDE with results for $h=0.0005$
($\langle R \rangle$, $\langle p_M \rangle$, $\langle u_M \rangle$),
and time intervals during which the system attains the TDE (see Fig.\ \ref{fig2} below);
these values for these steps are close.
For ``slower'' values of the frequency (for example, $\omega = 0.05$, $\chi = 0.1$ or $\omega = 0.01$, $\chi = 0.02$)
the trajectories were calculated using the combined scheme \cite{bib31},
where subsystem \eqref{eq3a} is solved by a more accurate Runge--Kutta 4-order method,
subsystem \eqref{eq3b} with a random term in right-hand side -- by the 2o2s1g method,
and the interaction of the subsystems is taken into account in a special way.
Using similar tests, the integration step $h=0.002$ and $\eps=10^{-4}$ were chosen.

We considered different $N$ and $T$, for each pair $(N,T)$
the set of 50--100 realizations was calculated.
When choosing $N$, $T$, we used the results \cite{bib21}:
for polyA fragments (SRP) the energy of the polaron decay is $N T_{crit} \approx 650$,
and for polyT chains (LRP) $N T_{crit} \approx 380$.

Though the main results were obtained with adopted ``accelerating'' values of the parameters,
we believe that the qualitative picture of the dynamics is similar in a wide range of the parameters for the systems \eqref{eq7}.
This assumption is confirmed by some test computations.
For example, in Figs.\ \ref{fig3} and \ref{fig4}
we plot the dynamics for 40-site chains with the parameters
$\{ \eta =0.456, \chi =1., \omega =0.5\}$ and $\{ \eta =0.456, \chi =0.2, \omega =0.1\}$
(SRP, $\eta$ value corresponds to polyA in quantum subsystem \eqref{eq3a}),
and in Fig.\ \ref{fig12} -- for LRP with different $\chi, \omega$ ($\eta = 2.4$ correspond to polyT).

\section{Results. Small Radius Polaron}
\label{sec4}

\begin{figure}[htb]
\includegraphics
[scale=0.25]{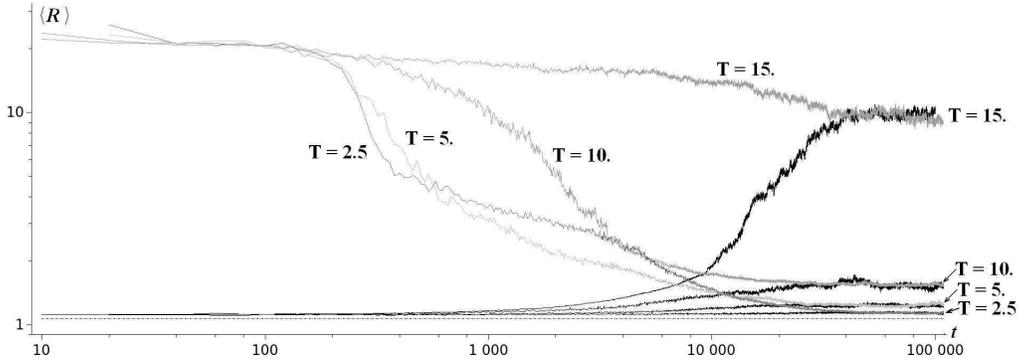}
\caption{$\langle R(t)\rangle$ from the polaron initial data (black) and
from uniform distribution $p_n(t=0) = 1/N$ (gray).
$N=40$, $\eta =0.456$, $\chi =1.$, $\omega =0.5$
($\chi / \omega = 2$), $\gamma = 4\omega$.
On the right, values of $T$ are indicated by arrows.
Dotted line $R = 1.07$ corresponds to polaron distribution $R(T=0)$.
\label{fig2}}
\end{figure}

For $V=0$ at small values $T$ polaron is immobile.
In the TDE the maximum of the probability $p_M$ decreases as $T$ grows,
however the polaron center $n_M$ does not change.
Fig.\ \ref{fig2} shows the $\langle R(t)\rangle$ curves (eq.\ \eqref{eqRv2}) for different $T$
in the chain of $N=40$ sites.
For each $T$ 2 sets of 50 realizations were calculated: from the polaron initial data
($\langle R(t=0)\rangle \approx 1.07$) and from uniform distribution $b_n(t=0) = 1/\sqrt{40}$
with random initial displacements $u_n$ and velocities $v_n$ from the Gauss distribution
for given $T$ ($\langle R(t=0)\rangle = 40$).

\begin{figure}[htb]
\includegraphics
[scale=0.34]{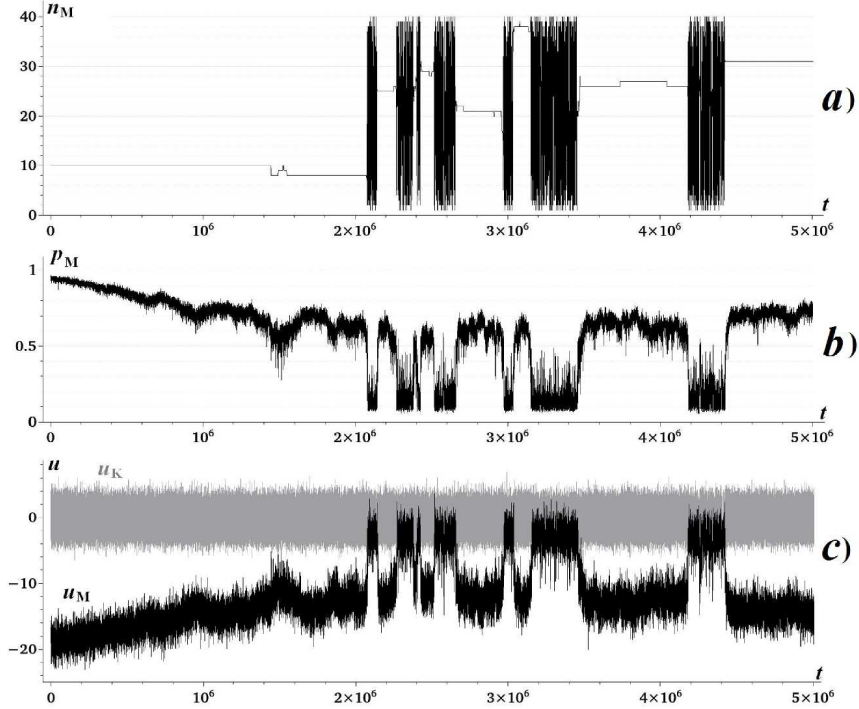}
\caption{Dynamics of one realization from the polaron initial data. $N=40$, $T=15$, $\eta =0.456$, $\chi =0.2$, $\omega =0.1$
($\chi / \omega = 2$), $\gamma = 3\omega$:
a) position of the probability maximum in the chain (site number) $n_M(t)$,
b) the value of $p_M(t)$,
c) displacement of the $n_M$-th site $u_M$ (black) and the largest displacement $u_K$ in the ``chain with window'' (gray).
\label{fig3}}
\end{figure}

For $T \approx T_{crit}$ (e.g., in Fig.\ \ref{fig2} for $N=40$, $T=15 \approx T_{crit}$,
$NT=600 < NT_{crit} = 650$),
the dynamics of a realization in the thermodynamic equilibrium state is as follows:
a polaron is formed at some site, exists for some time and then decays;
for some time the charge can be considered to be delocalized over the whole of the chain, then a polaron ``gathers'' at another site, etc.
Fig.\ \ref{fig3} shows the results of modeling of one realization for the chain of $N=40$ sites at $T=15$.
The initial stage of decaying of an ``ideal'' polaron is very large; for $t > 2\cdot 10^6$ the system oscillates near TDE in the switching mode:
for some time interval a charge is delocalized over the chain (e.g., from $t \approx 2.08\cdot 10^6$).
In this case the location of probability maximum $n_M$ ``jumps'' over the whole of the chain (Fig.\ \ref{fig3}\,a),
the value of the maximum is $p_M \approx 0.2$ (Fig.\ \ref{fig3}\,b).
On this interval the displacement of the site with the maximum probability $u_M$ is close to $u_K$ (Fig.\ \ref{fig3}\,c),
i.e.\ this is not a polaron state.
Then ($t \approx 2.15\cdot 10^6$) a charge is localized at one site ($n_M = 25$, Fig.\ \ref{fig3}\,a)
with a high probability $p_M \approx 0.6$ (Fig.\ \ref{fig3}\,b), besides $|u_M|> |u_K|$ (Fig.\ \ref{fig3}\,c),
i.e.\ a polaron ``gathers'' and exists for some time, then ($t \approx 2.3\cdot 10^6$) a charge again becomes delocalized, etc.
Note that for small $T$ in the TDE (Fig.\ \ref{fig2}),
$\langle R(T=5)\rangle \approx 1.24$ and $n_M (t) = const$ for each realization from polaron initial data;
$\langle R(T=10)\rangle \approx 1.55$, and here we observed in 3 realizations of 50 very short periods of delocalization.

\begin{figure}[htb]
\includegraphics
[scale=0.25]{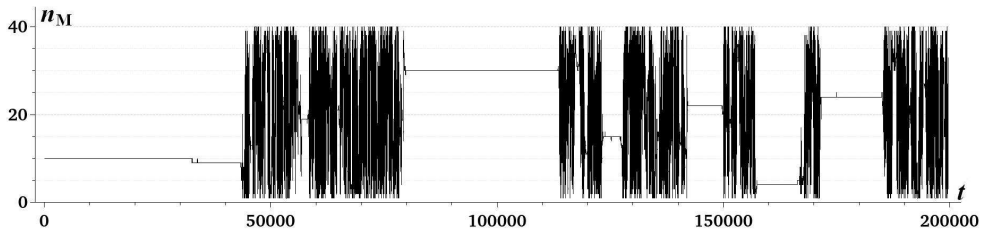}
\caption{Position of a site with the probability maximum $n_M(t)$.
One realization from the polaron initial data, $N=40$, $T=15$, $\eta =0.456$,
$\chi =1.$, $\omega =0.5$, $\gamma = 3\omega$.
\label{fig4}}
\end{figure}

\begin{figure}[htb]
\includegraphics
[scale=0.25]{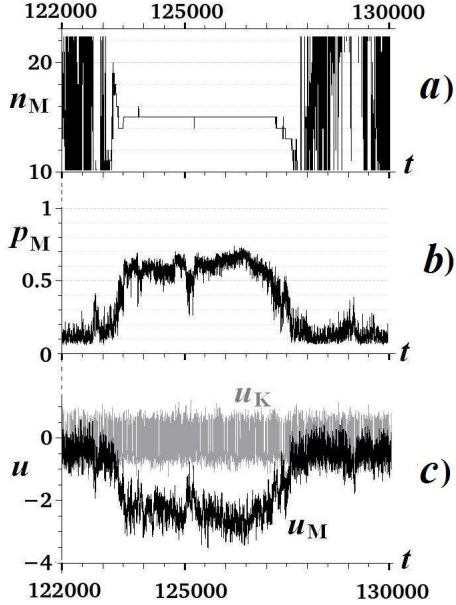}
\caption{A fragment of the dynamics from Fig.\ \ref{fig4} for the time $[122000, 130000]$:
a) the position $n_M$,
b) the value of $p_M$,
c) curves of $u_M$ (black) and $u_K$ (gray).
The plots are matched with respect to time.
\label{fig5}}
\end{figure}

The dynamics in Fig.\ \ref{fig3} is calculated for relatively slow values $\chi=0.2$, $\omega =0.1$,
for which initial polaron disruption occurs during the time $t > 10^6$.
Fig.\ \ref{fig4} shows a curve of the position $n_M(t)$ ($N=40$, $T=15$) for ``adopted'' values $\chi =1.$, $\omega =0.5$,
which were used in most of the computer simulations.
The initial polaron state disrupts during $t \sim 3\cdot10^4$.
Comparison of Figs.\ \ref{fig3}\,a and \ref{fig4} shows a qualitative similarity of the dynamics in the TDE --
the periods of time when a charge is delocalized over the whole of the chain
are changed by ``polaron'' intervals when a charge is localized at one site.
Notice that sometimes the picture demonstrates ``nearly polaron'' transitions to the neighboring site,
however the detailed consideration shows that the transition coincides with a delocalized state.
Fig.\ \ref{fig5} illustrates a fragment of the dynamics from Fig.\ \ref{fig4}
on the time interval $[122000,130000]$, where the jumps of the probability maximum to the neighboring site take place
(for example, at $t \approx 125000$). During the transition the probability maximum $p_M$ decreases and
the displacement of the site with the maximum is very close to displacements of other sites of the chain $u_M \approx u_K$,
i.e.\ the charge can be considered to be delocalized.
Regime in which a charge in the polaron state passes on from one site to another without delocalization intervals
was not observed.

\begin{figure}[htb]
\includegraphics
[scale=0.36]{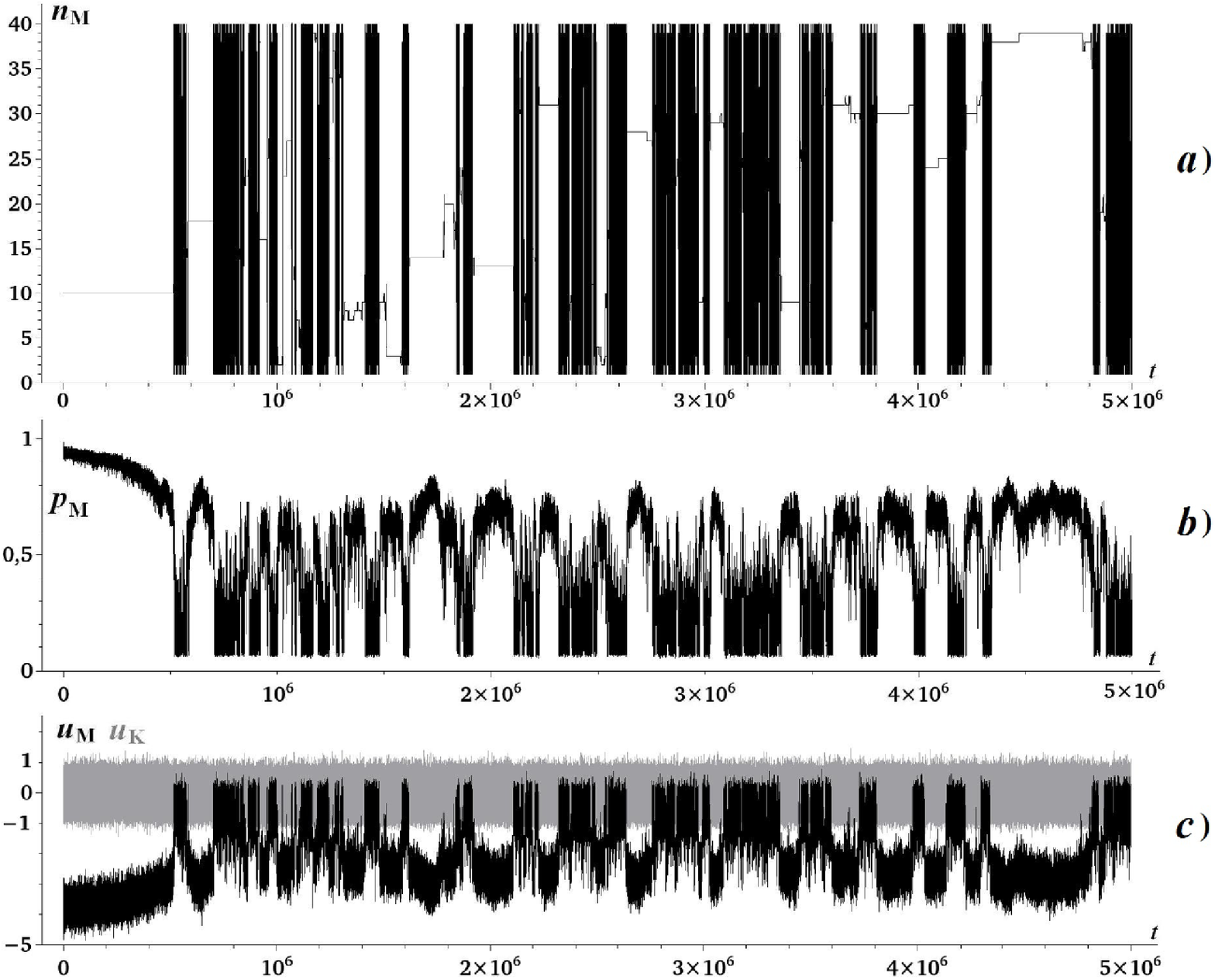}
\caption{Dynamics of one realization from the polaron initial data. $N=40$, $T=15$, $\eta =0.456$,
$\chi =1.$, $\omega =0.5$, $\gamma = 0.01$.
a) plot of $n_M(t)$,
b) plot of $p_M(t)$,
c) curves $u_M$ (black) and $u_K$ (gray).
\label{fig6}}
\end{figure}

The friction coefficient $\gamma$ plays important role in modeling.
To simulate the dynamics, we used rather large values of $\gamma > \omega$,
which accelerate the achievement of the TDE state.
For $\omega=0.5$ we tested small $\gamma$ values,
the picture looks similar in quality, although the initial polaron disrupts more slowly.
Fig.\ \ref{fig6} shows a realization for $\gamma=0.02\omega=0.01$;
the time interval of polaron disrupting $t \approx 5 \cdot 10^5$,
as for $\gamma=1.5$ (Fig.\ \ref{fig4}) initial polaron disrupts at $t \approx 4 \cdot 10^4$.

Note that the similar picture of initial polaron destroying
(Fig.\ \ref{fig3}\,b for $t<2.1\cdot 10^6$ and Fig.\ \ref{fig6}\,b for $t<5.5\cdot 10^5$) was
calculated
\cite{add2-09,add2-091} for the discrete nonlinear Schr\"{o}dinger equation,
which can be obtained as the limit of Holstein semiclassical model
when charge transfer between sites is very slow compared to the site oscillations: $\eta \ll \omega$.
We consider parameters $\eta \geq \omega$, and
the time interval for initial polaron destroying
at these parameters is much longer than in the works \cite{add2-09,add2-091}.

\begin{figure}[htb]
\includegraphics
[scale=0.2]{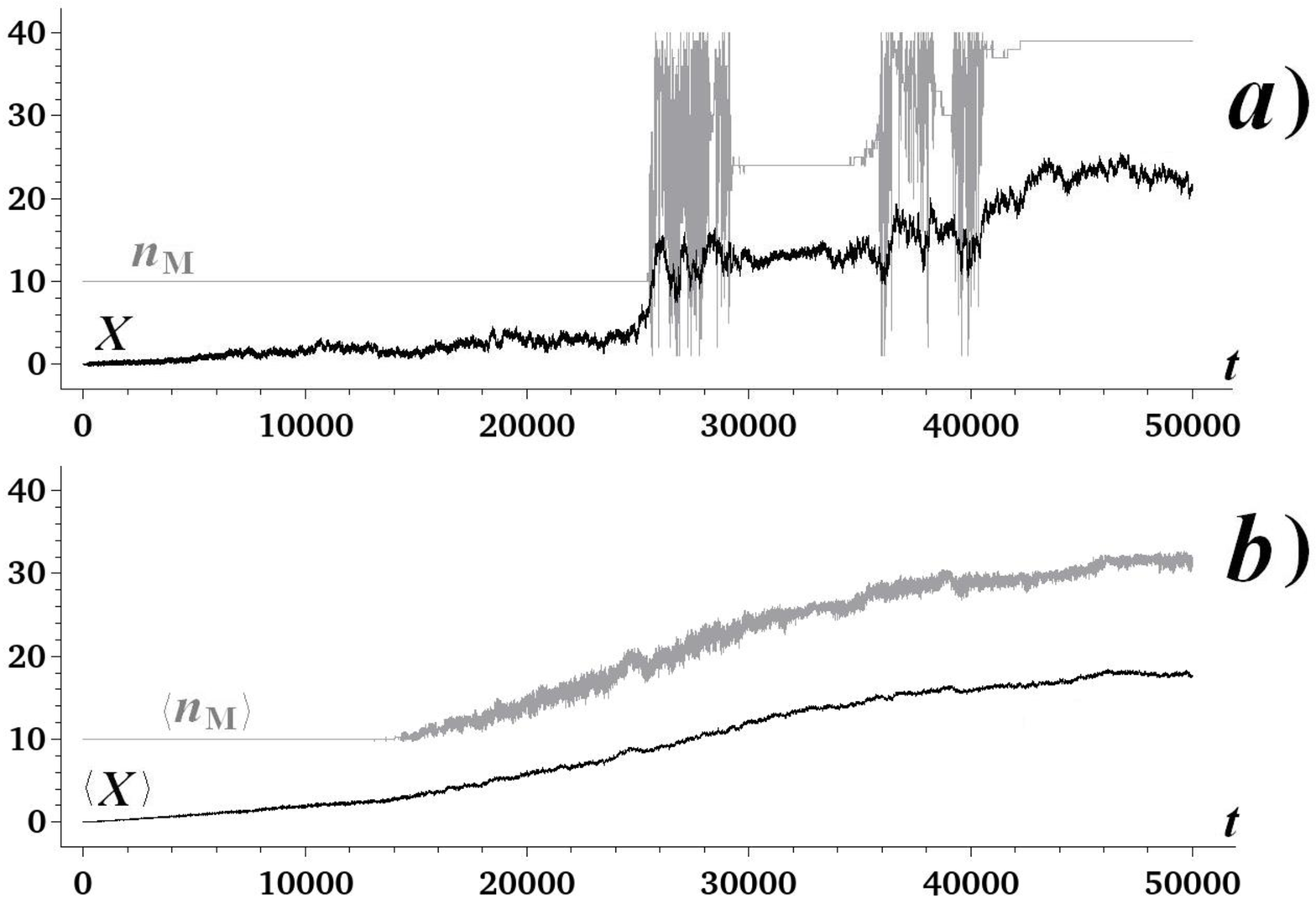}
\caption{SRP in the electric field with intensity $V=0.01$.
$N=40$, $T=15$, $\eta =0.456$, $\chi =1.$, $\omega =0.5$, $\gamma = 3\omega$.
a) The charge mass center $X(t)$ (black) and $n_M(t)$ (gray) for one realization,
b) $\langle X(t)\rangle$ (black) and $\langle n_M(t) \rangle$ (gray) averaged over 50 realizations.
\label{fig7}}
\end{figure}

The results of simulation in the chain under the action of the field with intensity $V \neq 0$
show that in realizations the center of the polaron state is immobile
(localized at one and the same site) until the polaron decays,
in this time interval $p_M(t)$ decreases, displacement $u_M$ at this site decreases in modulus,
but $n_M$ does not change (Fig.\ \ref{fig7}\,a for $t<25000$).
Then the charge passes on to the delocalized state and in this state migrates on an average
in the field direction. Then the probability maximum is localized in the region of the sites
with the lowest electron energy $\sim NV$.
When averaging over realizations, since disruption of the initial polaron occurs at different points in time,
more smooth curves are observed.  Fig.\ \ref{fig7} shows the curves
of the charge mass center $X(t)$ (Eq.\ \eqref{eq8}) and $n_M(t)$ calculated for the intensity $V=0.01$
for one realization (Fig.\ \ref{fig7}\,a), and $\langle X(t)\rangle$ and $\langle n_M(t)\rangle$
averaged over 50 trajectories (Fig.\ \ref{fig7}\,b).
The initial time interval $\langle X(t)\rangle$ (Fig.\ \ref{fig7}\,b) $0 < t < 14000$
corresponds to the stage of polaron decay.
The fragment $\langle X(t)\rangle$ for $16000 <t<36000$ fits well the straight line
and formally it can be used to estimate the average velocity of the charge $v$ (Eq.~\eqref{eq8a}).
However this is not the polaron mechanism of the transfer, as is seen from Fig.\ \ref{fig7}\,a.

Hence, the results of modeling demonstrate that for the SRP parameters in homogeneous chains a polaron
does not move
successively
from one site to another. Here we deal with
a switching mode between immobile polaron state and the delocalized one,
and the transfer takes place in the delocalized state.

\section{Results. Large Radius Polaron}
\label{sec5}

\begin{figure}[htb]
\includegraphics
[scale=0.2]{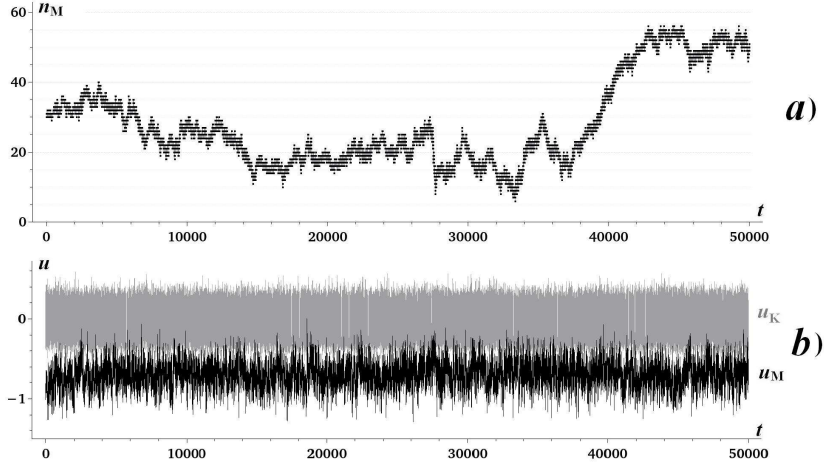}
\caption{Large radius polaron (polyT fragment), $\eta =2.4$, $\chi =1.$, $\omega =0.5$, $\gamma = 3\omega$.
$N=60$, $T=3$ ($NT < NT_{crit} \approx 380$), one realization, at $t=0$ polaron with the center on the site $n_0=30$.
a) Curve of $n_M(t)$,
b) plots $u_M$ (black) and $u_K$ (gray).
\label{fig8}}
\end{figure}

Since for $T=0$ the highest probability (in the center of the polaron) is rather small ($p_M \approx0.22$, Fig.\ \ref{fig1})
and, accordingly, the polaron displacements of the sites are small (Eq.\ \eqref{eq6}),
the criterion of ``polaron displacements'' (which means that
the greatest in modulus displacement will be of the site with the polaron center)
is not always fulfilled here.
Fig.\ \ref{fig8} shows the time dependencies of the $n_M$ (Fig.\ \ref{fig8}\,a) and $u_M$ (Fig.\ \ref{fig8}\,b).
It is seen that the polaron moves successively, however the displacements $u_M$ sometimes
can be closer to zero than $u_K$ (Fig.\ \ref{fig8}\,b).

As distinct from SRP, here the region of localized charge
displaces successively from one site to another,
i.e.\ LRP can moves along the chain.
Therefore in this case we study time dependencies averaged over realizations.

In homogeneous chains disruption of polaron states depends on the heat energy of a chain $NT$ \cite{bib20,bib21}.
At the same temperature $T$ in short chains ($NT < N T_{crit}$) a polaron exists
while in long ones ($NT > NT_{crit}$) it decays.
For the chains of different lengths, polaron decays with different rates.
However, if at $t=0$ there is polaron in a chain, then for some time it exists.
We investigate how polaron moves in chains of different lengths $N$
under the action of a field with constant intensity $V$.

\begin{figure}[htbp]
\includegraphics
[scale=0.2]{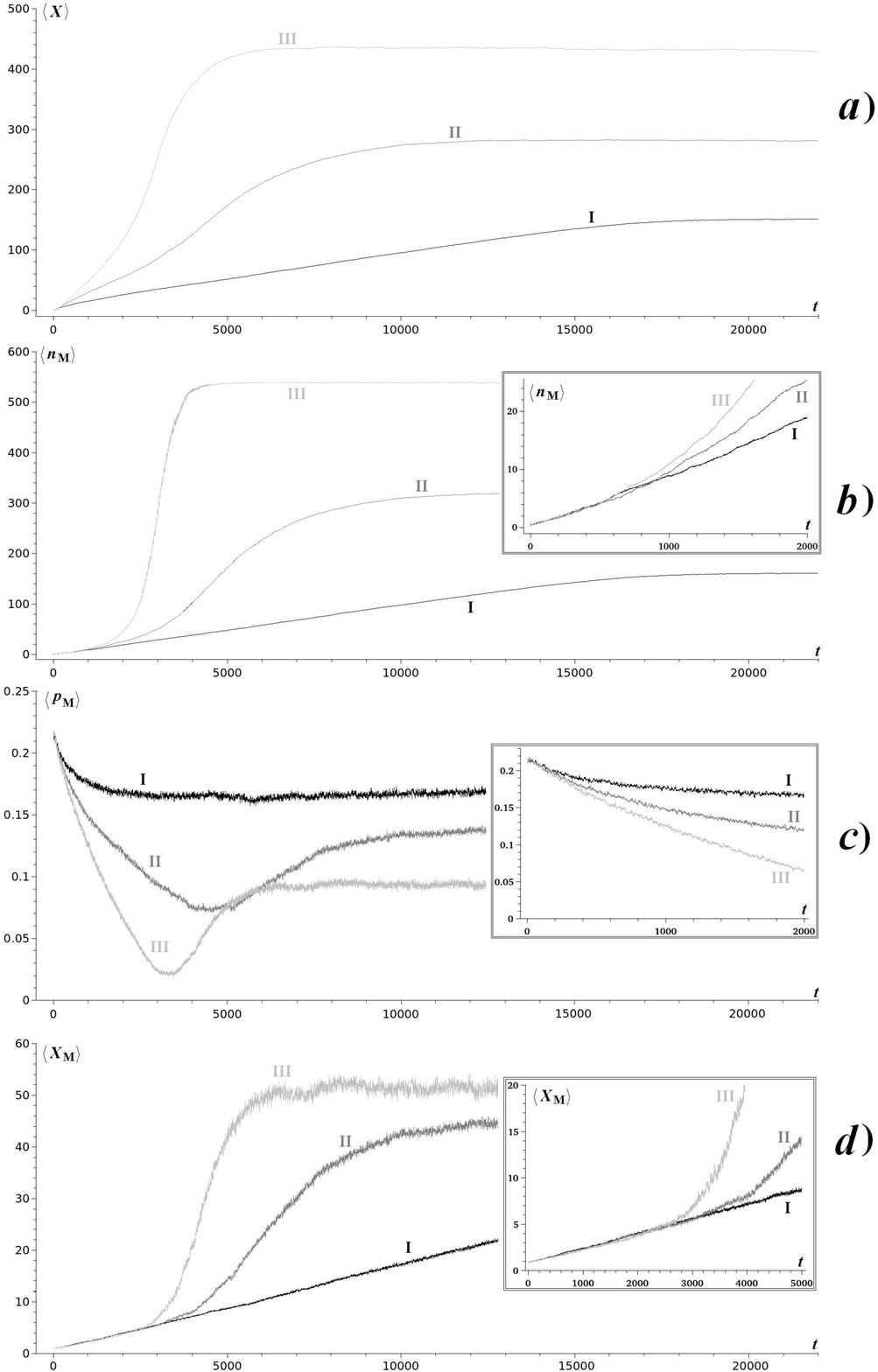}
\caption{LRP, $\eta =2.4$, $\chi =1.$, $\omega =0.5$, $\gamma = \omega$, $T=1$, $V=-0.001$;
chains of 200 sites (I, black), 400 sites (II, gray) and 700 sites (III, light gray).
Each curve is mean over 100 realizations. The inserts demonstrate the zoomed initial stage.
Initial polarons have center $n_0$ at the 20-th site for $N=200$, at the 70-th site for $N=400$,
and at the 150-th site for $N=700$.
\label{fig9}}
\end{figure}

The method of estimating the charge mobility $\mu$ for different $V$
(Eqs.\ \eqref{eq8},\eqref{eq8a}) fails, since in the chains of different lengths
the initial polaron decays differently.
Fig.\ \ref{fig9} shows the mean displacement of the charge center of mass $\langle X(t) \rangle$  (a),
positions of the maximum  $\langle n_M(t) \rangle$  (b), and the maximum probability $\langle p_M(t) \rangle$ (c)
for the chains of different lengths in the field of intensity $|V|=0.001$ for $T=1$
(the sign of $V$ determines the direction of movement).
For polyT parameters, the critical value is $NT_{crit} \approx 380$,
i.e.\ for $T=1$ in the chains consisting of 200 sites a polaron exists,
$N=400$ is near-border length of a chain
and for the chain of 700 sites, a polaron decays.

Fig.\ \ref{fig9}\,(a--c) shows that $\langle X(t) \rangle$, $\langle n_M(t) \rangle$ and  $\langle p_M(t) \rangle$
rather quickly diverge (at $t < 500$).
Curves III (for $N=700$) show that initial polaron moves gradually  disrupting,
and after $t \approx 2000$ velocity of $\langle n_M(t) \rangle$ increases greatly,
and slope of $\langle X(t) \rangle$ changes. On this time polaron disrupts,
and then waves of $b_n$ move along the field direction.
At $t \approx 4000$ $\langle p_M(t) \rangle$ reaches the end of the chain
(after that $\langle n_M(t) \rangle$ does not change), and then
the probability maximum is localized in the region of the sites
with the lowest electron energy $\sim NV$ and gradually $p_M(t)$ increases.
And after $t \approx 6000$ this system attains new thermodynamic equilibrium state
($\langle X(t>6000) \rangle \approx const$, as well as $\langle n_M(t) \rangle$ and $\langle p_M(t) \rangle$).
For 400-site chain (curves II), the picture is qualitatively similar, but more extended in time.
And for $N=200$ (curves I) at $t \approx 2000$ polaron is formed corresponding to this conditions
($T$ and $V$), and polaron moves along the chain; at $t \approx 18000$ it reaches the end of the chain
(after that $\langle X(t) \rangle \approx const$, $\langle n_M(t) \rangle \approx const$, Fig.\ \ref{fig9}\,a),b).

\begin{figure}[htb]
\includegraphics
[scale=0.3]{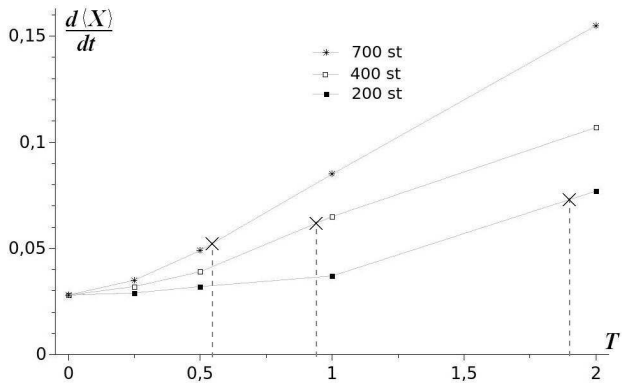}
\caption{The slope of $\langle X(t) \rangle$ in the initial interval (as long as polaron exists)
at $T=0.25,0.5,1,2$ for chains of $N=200$ sites (black squares),
$N=400$ (open squares), $N=700$ (stars).
LRP, $\eta =2.4$, $\chi =1.$, $\omega = \gamma = 0.5$, $V=0.005$.
The crosses indicate the conditional boundary $NT=380$
between the polaron and delocalized states in the TDE at a given chain length.
\label{fig10}}
\end{figure}

The slope of $\langle X(t) \rangle$ in the initial interval (as long as polaron exists) is different
for different $N$; e.g., in Fig.\ \ref{fig9}\,a) these intervals are:
$t<1400$ for $N=700$, $t<2700$ for $N=400$, and $t<18000$ for $N=200$.
As shown in Fig.\ \ref{fig10}, the difference decreases with decreasing temperature
and for $T=0$ $dX(t)/dt$ is the same for all chains with length $N \gg R(T=0)$.

Instead of $\langle X(t) \rangle$  we calculated truncated variant $\langle X_M(t) \rangle$ (Eq.\ \eqref{eq9})
where only the motions of the polaron top are used, see Fig.\ \ref{fig9}\,d.
Calculations demonstrate that for chains of different lengths, the curves $\langle X_M(t) \rangle$
coincide on a far greater time interval, than $\langle X(t) \rangle$, $\langle n_M(t) \rangle$ and $\langle p_M(t) \rangle$:
in Fig.\ \ref{fig9}\,d the common segment $t > 2500$, separation of the other plots becomes noticeable at a far lesser time.
Notice that for a 200-site chain (black line (I) in Fig.\ \ref{fig9}),
in which a polaron does not decay in TDE at $T=1$, $\langle X_M(t) \rangle$ behaves
like a straight line until a polaron reaches the end of chain. For still longer chains,
a polaron decays at some moment of time which is evident from a sharp change in $\langle X_M(t) \rangle$
(for a 400-site chain (II) $t \approx 4000$, and for a 700-site chain (III) $t \approx 2500$).

By analogy with the estimation of the charge mobility \eqref{eq8},\eqref{eq8a} for intensity $V\neq 0$,
based on $\langle X_M(t) \rangle = v_M t$
we estimate the ``mean velocity of the polaron'' $v_M$ for different $V$ and
get a ``polaron mobility'' $\mu_M$ by substituting $v_M = \mu_M V$.

Table \ref{tab1} lists the results of calculations of $v_M$ based on the slope
of the linear fragment $\langle X_M(t) \rangle$ averaging over 100 realizations,
common for the chains of different lengths.
The simulations are performed for different $T$, the parameter values are $\eta =2.4$,
$\chi =1.$, $\omega = 0.5$, $\gamma = 0.3$.
It is evident that the change in $V$ is proportional to the change in $v_M$.
Though the values of $T$ grow twice, the calculated values of $\mu_M$ are very close
and therefore one cannot single out the dependence $\mu_M(T)$.
The dependence of $\mu_M (T)$ for small $T$ can be more exactly
due to quadratic increase in the number of realizations which is computationally expensive.
The calculations carried out suggest that the polaron mobility for $T\to 0$ is very small,
substitution of the DNA parameter values leads to $\tilde{\mu}  = (ea^2/\hbar )\mu_M \approx 0.005$\,cm$^2/$(V$\cdot$sec).
Notice that for $T=0$ (Eq.\ \eqref{eq8},\eqref{eq8a}) the charge mobility is $\mu_0 \approx 9.2$,
and the polaron mobility $\mu_{M0} \approx 2.0 = \mu_0 p_M$
(since for $T=0$ the polaron velocity measured from $n_M(t)$ coincides with $\Delta X/\Delta t$)
is close to the values of Table~\ref{tab1}.

\begin{table}
\caption{Polaron velocity $v_M$ and mobility $\mu_M$
\label{tab1}}
\begin{center}
\begin{tabular}{c|c|c|c|c}
$T$ & $v_M$, $N=200$ & $v_M$, $N=400$ & $v_M$, $N=700$ & $\mu_M = v_M / V$ \\
 & $V=0.0005$ & $V=0.001$ & $V=0.005$ & \\
\hline
0.25 & 0.0012 & 0.0023 & 0.011 & 2.2-2.4 \\
0.5  & 0.0013 & 0.0023 & 0.012 & 2.3-2.6 \\
1.   & 0.0012 & 0.0024 & 0.012 & 2.4 \\
\hline
\end{tabular}
\end{center}
\end{table}

\begin{figure}[htb]
\includegraphics
[scale=0.25]{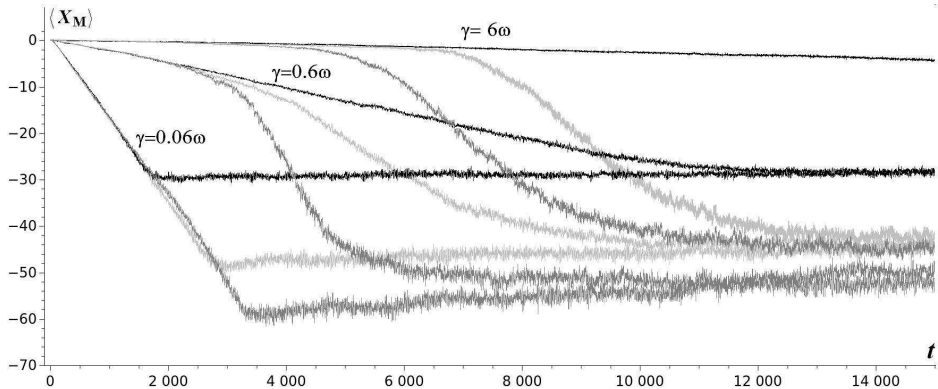}
\caption{$\langle X_M(t)\rangle$ for $N=200$ (black curves), $N=400$ (light gray)
and $N=700$ (gray). $T=1$, $\eta =2.4$, $\chi =1.$, $\omega =0.5$, $V=0.001$
(initial polarons are localized near right end of the chain, the center of the polaron is
at the same distance from the right end as in Fig.~\ref{fig9});
different $\gamma$ values.
\label{fig11}}
\end{figure}

For LRP the value of $\gamma$ is significant.
At $T=0$ for fixed $V$ velocity of polaron depends on $\gamma$ \cite{bib7},
the smaller $\gamma$, the greater velocity.
Fig.\ \ref{fig11} shows $\langle X_M(t)\rangle$ for different $\gamma$ at $T=1$.
For $\gamma=3$ slope of $\langle X_M(t)\rangle$ is small, and for chains $N=400$ and $N=700$
the initial polaron has time to disrupt.
For $\gamma=0.03$ slope of $\langle X_M(t)\rangle$ is large
and for all chains polaron manages to reach the end of the chains.

\begin{figure}[htb]
\includegraphics
[scale=0.45]{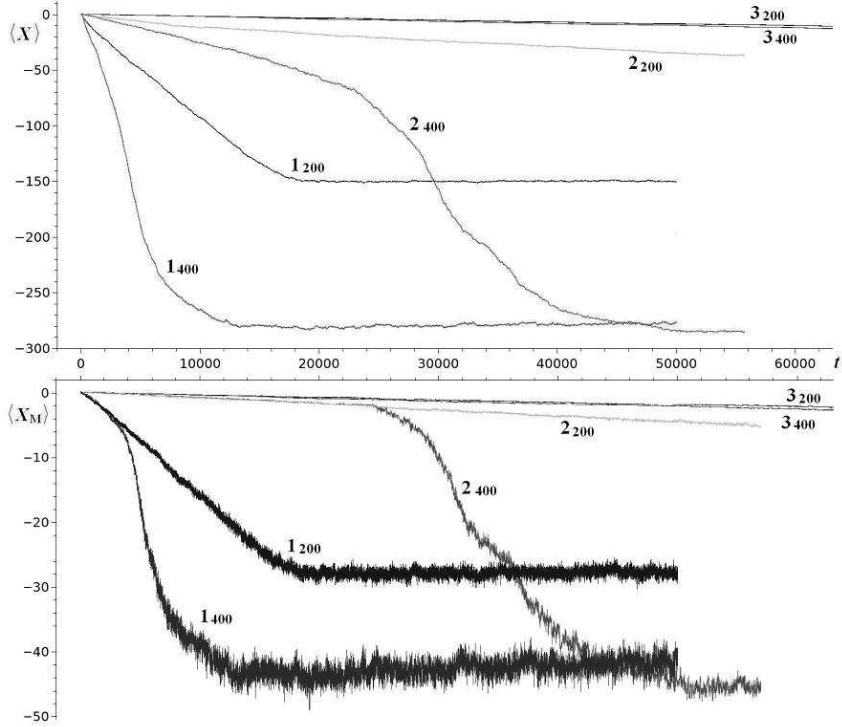}
\caption{Curves $\langle X(t)\rangle$ (top) and $\langle X_M(t)\rangle$ (bottom)
for chains of 200 and 400 sites (subscript denotes chain length). $T=1$, $\eta =2.4$, $V=0.001$,
$\gamma=\omega$, initial polarons are localized as in Fig.\ \ref{fig9}.
Parameters for curves (1): $\chi =1.$, $\omega =0.5$;
for (2): $\chi =0.2$, $\omega =0.1$;
for (3): $\chi =0.02$, $\omega =0.01$.
\label{fig12}}
\end{figure}

Also the ratio $\eta/\omega$
is important for demonstration of chain length effects.
Fig.\ \ref{fig12} shows $\langle X(t)\rangle$ and $\langle X_M(t)\rangle$
of dynamics of the initial polaron (averaged over 24 realizations) for $\eta=2.4$ and different values of $\chi,\omega$
with the same ratio $\chi/\omega =2$ (subscripts in Fig.\ \ref{fig12} denote chain length).
Although in the TDE state the mean values should be the same, but the time to attain it may be very long.
For curves 3 in Fig.\ \ref{fig12} with ($\chi=0.02,\omega=0.01$) on this time of integration
there is no noticeable difference in the curves for $N=200$ ($3_{200}$) and $N=400$ ($3_{400}$)
for $\langle X(t)\rangle$ and $\langle X_M(t)\rangle$.
Note that for ``realistic'' DNA parameter values in Holstein model \cite{bib24,bib25},
$\eta/\omega$ is of the order of hundreds, and results are similar to curves 3.

\section{Discussion and conclusions}

Using a computational experiment we studied the dynamics of a polaron in a chain with Langevin thermostat
under the action of a constant electric field. We considered the cases of a small radius polaron and a large radius polaron.

It is shown that
for the parameters corresponding to small radius polaron, the polaron is immobile.
Calculations of dynamics demonstrate that polaron exists
for some time and then decays;
for some time the charge is delocalized over the whole chain,
then a polaron ``gathers'' at another random site.
Charge transfer takes place in the delocalized state,
i.e., for the SRP parameters, a charge does not move along the chain like a polaron.
In this work we consider possibility of charge transfer by polaron mechanism;
a number of interesting questions, e.g.\
mean ratio of time intervals ``polaron/delocalized state'' in the TDE, or
how time of initial polaron destroying depends on $N$ and $T$ will be studied later.

Theoretical models of SRP motion imply that a polaron under the action of temperature fluctuations hops from one site to the neighboring one
or to one of the neighboring sites within a small region \cite{bib19,bib32,add02-10,add02-11}.
We calculated the dynamics in the chains consisting of 40 to 200 sites.
The results demonstrate that a charge in the delocalized state is ``spread'' over the whole chain
rather than over some of its regions with the center of the former polaron,
and the position of the new polaron which gathered after delocalization interval is not related to the position of the previous polaron.
It looks like intermittency in dynamical systems \cite{add02-12}.

This behavior depends on the polaron radius $R$; for polyC chains ($R \approx 1.2$, matrix element $\eta =0.623$, $\nu = 0.041$\,eV \cite{bib24,bib25})
on some time intervals there is a consistent polaron movement, but such intervals are separated by intervals of delocalization
along the whole chain; for polyA ($R \approx 1.07$) and Holstein SRP ($R \approx 1.02$, parameters from footnote 17 of \cite{bib19})
such consistent polaron motion was not observed.

Perhaps, at very small $T$, this regime passes into the hopping mechanism considered by Holstein \cite{bib19} or variable range hopping \cite{bib32}.
Directly during simulations, we did not observe such regimes (and we do not know the papers in which the authors observed such regimes in direct simulations),
possibly because theoretically in this case the charge is localized at one site for a very long time.

Large radius polaron can move over the chain.
In the semiclassical Holstein model the region of existence of polaron in thermodynamic equilibrium state
depends not only on temperature but also on the chain length. Therefore in modeling from initial polaron data
the slope of $ \langle X(t) \rangle  = \langle \sum_n p_n (n - n_0)  \rangle$
differs for the chains of different lengths at the same temperature.
We showed numerically that for the chains of different lengths, the ``polaron center mass''
$ \langle X_M(t) \rangle  = \langle p_M (n_M - n_0)  \rangle$
behaves similarly until the polaron decays (for the same $T$ and intensity $V$ of electric field).
A similar slope $\langle X_M(t) \rangle$ enables one
to find the mean polaron velocity and, by analogy with the charge mobility,
estimate the polaron mobility $\mu_M$.
The calculated values of $\mu_M$ for small $T$ are close to the value at $T=0$,
which does not contradict the assumption $ \mu_M (T \to 0) = \mu_M (0) $.
The calculations performed demonstrate that at zero temperature the polaron mobility is small but nonzero.
These results agree with earlier obtained data on the Holstein polaron motion along an unperturbed chain (for $T=0$) \cite{bib7,bib9}.

We considered a simple model, classical chain without dispersion (in \cite{bib33} dispersion is used to model stacking interaction of the DNA strand).
More detailed Peyrard--Bishop model with nonlinear interaction
between neighboring base pairs \cite{bib34,bib35}
for the case of small site displacements can be reduced
to the form similar to that of the dispersion term in equations for crystals.
The values of the charge mobility in DNA obtained in some works on the basis of different models are very small \cite{bib36,bib37,bib38}.
For this reason different mechanisms of charge transfer with the use of nonlinear excitations are studied \cite{bib33,bib40,bib41}.
It has been found that a metastable quasi-particle may be transported at a distance up to 200 base pair \cite{bib42,bib43} without an external electric field.
Such mechanism may be considered as an alternative one to the polaron mechanism.

According to the results of modeling, for the LRP parameter values
the charge transfers faster in the delocalized state.
Thus, in Fig.\ \ref{fig9} it is shown that in 700-site chain the charge maximum $n_M$ reaches the end at the time $t\approx 5000$.
The same time interval is taken to ``gather'' the probability at the end of a chain, the greatest value is
$\langle p_M(t = 10000) \rangle \approx 0.09$.
While for the 200-site chain polaron passes it during $t \approx 20000$, but the probability maximum is always
$\langle p_M \rangle > 0.15$.
Which regime is better
(and can these modes be realized in experiments)
-- it is the problem for further investigations.

The ratio $\eta/\omega$ is important for observing chain length effects, such as
difference in $\langle X(t)\rangle$, $\langle p_M(t)\rangle$, $\langle n_M(t)\rangle$ for different $N$.
The larger this ratio, the smaller the difference in the slopes of $\langle X(t)\rangle$ at the same $V$.

We consider the chains with free ends. For $V=0$ we made several numerical tests
(for SRP $N=40$, $T=10$, $T=15$, and for LRP $N=200$ and $N=700$, $T=1$)
in circle chain (i.e., periodic boundary conditions).
The test results (averaged values) are similar.
For $V \neq 0$ this question needs further research.

We carried out numerical simulations  for some sets of parameter values,
with $\chi \approx \omega$, $\eta \geq \omega$. Based on this extensive work, we can expect that qualitatively similar pattern of the polaron dynamics exists in a wide range of parameters.
Probably, direct simulations in other mixed quantum-classical models with Langevin term \cite{bib15,addR1,add022-2,add022-3}
might demonstrate some similar features of polaron dynamics, different dynamics at the same temperature in the chains with different length,
such as dependence of $\langle p_M(t)\rangle$ or $\langle X(t)\rangle$ on the chain length.


\section*{Acknowledgments}

We are grateful to the Keldysh Institute of Applied Mathematics of the Russian Academy of Sciences
for providing high-performance computational facilities of k-60 and k-100.
This work was supported in part by Russian Foundation for Basic Research, grants 19-07-00406 and 17-07-00801,
and the Russian Science Foundation, project 16-11-10163.

\bibliography{NFbib4}

\end{document}